CORRESPONDENCE

To the Editors of 'The Observatory'

*The sunspot observations by Toaldo and Comparetti at November 1779*

The recovery of old sunspot observations is critical to reconstruct the solar activity in the past and understand the long-term solar variability[1-2]. Some works have been focused on this task[3-4], collecting large amounts of daily counts of sunspot group numbers. The number of days with records per year in these collections is very variable from 1610 to the present. There is, at least, one observation per day from 1848 to present, but there is a low frequency of observations in some years of the 17$^{th}$ and 18$^{th}$ century. In particular, the interval 1777-1795 is one of the periods with scarce observations in the database[4].

We report here a sunspot observation carried out the 3$^{rd}$ November 1779 by Giuseppe Toaldo and Andrea Comparetti. Andrea Comparetti was professor of Medicine at Padova University and Giuseppe Toaldo was the director of the astronomical observatory of this city. In a dissertation about the severe droughts during the winter of 1779 Toaldo wrote[5]: "*Un' osservazione singolare, degna da ponderarsi in quest' anno, fu l'infinità di macchie che si obsservarono nel sole, sempre, ma segnatamente nell' Inverno continuano tuttavia, o piuttosto ripigliano e risorgono anche oggi 3 Novembre col Sig. Dottor Compareti, dotto Fisico non meno che valente Medico, ne abbiamo contate almeno 17 in varj ammassi, e alcuna di esse aveva il diametro più grande di quello della terra, poichè certamente più d' un minuto era la loro apparente grandezza*". [One special observation, worthy to take into account in this year, was the huge number of sunspots observed in the sun, continually, but particularly in winter still continuing, or rather recovering and rising again even today 3$^{rd}$ November with the Dr. Comparetti, erudite physician but not less talented medic, we have counted at least 17 in various groups, and some of them had the diameter greater than the earth diameter, since the apparent size was greater than a minute.]

This observation is particularly interesting because nobody observed the solar disc on 3$^{rd}$ November 1779 and neither in close dates. The previous available observation is on 22$^{nd}$ October 1779 and the next one was made on the 14$^{th}$ January 1780 (both by Johann

Casper Staudacher in Nuremberg)[6-7]. Note that Toaldo was an active observer of aurorae[8]. Moreover, it is possible that Toaldo or Comparatti observed sunspots frequently. The localization of relevant observations made by Toaldo or Comparatti would be of great interest for this kind of studies.

We apreciate the support of EU, Junta de Extremadura, and Ministry of Economy and Competitiveness (consortium IMDROFLOOD, Research Group Grant GR15137, IB16127, and AYA2014-57556-P).

Yours faithfully,

FERNANDO DOMÍNGUEZ-CASTRO

Instituto Pirenaico de Ecología

Consejo Superior de Investigaciones Científicas (IPE-CSIC)

Avda. Montañana, 1005.

E-50059 Zaragoza, Spain

JOSÉ M. VAQUERO

Departamento de Física

Centro Universitario de Mérida, Universidad de Extremadura

Avda. Santa Teresa de Jornet, 38

E-06800 Mérida, Badajoz, Spain

2017 April 26